\documentclass[showpacs,aps,twocolumn,superscriptaddress,prb,amsmath]
{revtex4}
\usepackage{txfonts}
\usepackage{color}
\usepackage[latin9]{inputenc}
\usepackage{amssymb}
\usepackage{graphicx}
\usepackage{dcolumn}
\usepackage{bm}
\usepackage{hyperref}
\usepackage{color}

\begin{document}
\title{Ground state configuration of hydrogenated Biphenylene sheet: structure, stabilities, electronic and mechanical properties from first-principles calculations}
\author{Yujie Liao}
\affiliation{Hunan Key Laboratory of Micro-Nano Energy Materials and Devices, Xiangtan University, Hunan 411105, P. R. China}
\affiliation{Laboratory for Quantum Engineering and Micro-Nano Energy Technology and School of Physics and Optoelectronics, Xiangtan University, Hunan 411105, P. R. China}

\author{Xizhi Shi}
\affiliation{Hunan Key Laboratory of Micro-Nano Energy Materials and Devices, Xiangtan University, Hunan 411105, P. R. China}
\affiliation{Laboratory for Quantum Engineering and Micro-Nano Energy Technology and School of Physics and Optoelectronics, Xiangtan University, Hunan 411105, P. R. China}

\author{Tao Ouyang}
\affiliation{Hunan Key Laboratory of Micro-Nano Energy Materials and Devices, Xiangtan University, Hunan 411105, P. R. China}
\affiliation{Laboratory for Quantum Engineering and Micro-Nano Energy Technology and School of Physics and Optoelectronics, Xiangtan University, Hunan 411105, P. R. China}

\author{Jin Li}
\email{lijin@xtu.edu.cn}
\affiliation{Hunan Key Laboratory of Micro-Nano Energy Materials and Devices, Xiangtan University, Hunan 411105, P. R. China}
\affiliation{Laboratory for Quantum Engineering and Micro-Nano Energy Technology and School of Physics and Optoelectronics, Xiangtan University, Hunan 411105, P. R. China}

\author{Chunxiao Zhang}
\affiliation{Hunan Key Laboratory of Micro-Nano Energy Materials and Devices, Xiangtan University, Hunan 411105, P. R. China}
\affiliation{Laboratory for Quantum Engineering and Micro-Nano Energy Technology and School of Physics and Optoelectronics, Xiangtan University, Hunan 411105, P. R. China}

\author{Chao Tang}
\affiliation{Hunan Key Laboratory of Micro-Nano Energy Materials and Devices, Xiangtan University, Hunan 411105, P. R. China}
\affiliation{Laboratory for Quantum Engineering and Micro-Nano Energy Technology and School of Physics and Optoelectronics, Xiangtan University, Hunan 411105, P. R. China}

\author{Chaoyu He}
\email{hechaoyu@xtu.edu.cn}
\affiliation{Hunan Key Laboratory of Micro-Nano Energy Materials and Devices, Xiangtan University, Hunan 411105, P. R. China}
\affiliation{Laboratory for Quantum Engineering and Micro-Nano Energy Technology and School of Physics and Optoelectronics, Xiangtan University, Hunan 411105, P. R. China}

\author{Jianxin Zhong}
\affiliation{Hunan Key Laboratory of Micro-Nano Energy Materials and Devices, Xiangtan University, Hunan 411105, P. R. China}
\affiliation{Laboratory for Quantum Engineering and Micro-Nano Energy Technology and School of Physics and Optoelectronics, Xiangtan University, Hunan 411105, P. R. China}

\begin{abstract}
Based on first-principles calculations, the ground state configuration (Cmma-CH) of hydrogenated Biphenylene sheet (Science, 372, 852, 2021) is carefully identified from hundreds of possible candidates generated by RG2 code (Phys. Rev. B., 97, 014104, 2018). Cmma-CH contains four benzene molecules in its crystalline cell and all of them are inequivalent due to its Cmma symmetry. The hydrogen atoms in Cmma-CH bond to carbon atoms in each benzene with a boat-like (boat-1:DDUDDU) up/down sequence and reversed boat-1 (UUDUUD) sequence in adjacent benzene rings. It is energetically less stable than the previously proposed allotropes (chair, tricycle, stirrup, boat-1, boat-2 and twist-boat) of hydrogenated graphene, but its formation energy from hydrogenating Biphenylene sheet is remarkably lower than those for hydrogenating graphene to graphane. Our results confirm that Cmma-CH is mechanically and dynamically stable 2D hydrocarbon phase which is expectable to be experimentally realized by hydrogenating the synthesized Biphenylene sheet. The HSE06 based band structures show that Cmma-CH is an indirect band gap insulators with a gap of 4.645 eV.
\end{abstract}

\maketitle
\indent It is always an exciting thing to experimentally realize the theoretically predicted new crystals \cite{NaCl, 2DB, BH} but it is not always happen. Large numbers of new materials\cite{calypso1, NRM, PRL_N1, PRL_N2, PF_PRL, LJjpcl, YPL, oyPSS} have been predicted by first-principles calculations in the past decades and just few of them have been experimentally synthesized \cite{exp1, PRLexpN, expN, PRL_exp2}. In the two-dimensional (2D) case, graphdiynes, Phagraphene, phosphorene and graphane are some successful stories. The sp-sp$^2$ hybridized graphdiynes were previously predicted in 1987 \cite{yne1987} and some of them have been realized in recent years \cite{yne1, yne2, yne3}. The low-energy Phagraphene with Dirac-Cone semi-metallic property was proposed in 2015 \cite{Pha}, and it has been experimentally synthesized in 2019 \cite{jacsPha}. The chair-type phosphorene (blue phosphorene) was theoretically predicted in 2014 \cite{PRL_zhu} and it has been successfully grown on Au(111) surface by molecular beam epitaxy method in 2016 \cite{NL2016}. The ground state configuration of hydrogenated graphene (chair-type graphane) predicted in 2003 \cite{prb2003} has been successfully synthesized \cite{S2009} in 2009 by hydrogenating graphene \cite{gra2004}.

Very recently, the previously predicted Biphenylene sheet \cite{netc2010}, a graphene allotrope with metallic property, was just synthesized \cite{netc2021} through a two-step interpolymer dehydrofluorination polymerization approach (called as Biphenylene network). After we have studied this work, our research interests are attracted by such an experimentally realized new 2D carbon crystal although it is just an useless normal metallic phase. It is because that such a new phase provides us the opportunity to create a new 2D hydrocarbon by the same way of synthesizing graphane (Graphene+H$_2$) \cite{S2009}. In this work, we try to fix a series of physical problems by means of first-principles calculations to find out some useful data for guiding experimental researchers to explore such a new potential 2D hydrocarbon. For example, how many possible configurations of hydrogenated Biphenylene sheets we can have and how about their formation energies? Which is the ground state configuration and how about its phase stabilities? How hydrogenation affects on the mechanical and electronic properties of the Biphenylene sheet?

To generate enough structure configurations of hydrogenated Biphenylene sheet, its 2$\times$2 supercell (containing 24 carbon atoms) is considered as the start point. Hydrogen atoms are randomly bonded to carbon atoms in the supercell through our previously developed RG$^2$ code \cite{Shi18, hcyprl, yhcprb}, which is designed for generating crystal structures with well-defined structural feature \cite{gzhprb, wlNC, gdlNE, splpb, jnpss, lzqass, znprb, yxjap}. For each carbon atom in the supercell, there are two different hydrogenating manners, namely, up (U) and down (D). The configuration space can be easily evaluated as 2$^{24}$=16777216, but most of them are distorted with high-energy and some of them are equivalent due to symmetry. It is difficult to visit all the 16777216 configurations to reject the distorted and equivalent ones. We just use RG2 to randomly visit 100000 of them and find 173 possible candidates after excluding the reduplicate ones and the distorted ones containing U:UUU or D:DDD bond-lists (carbon atom bonding to three neighbours with same up/down sequence).

The 173 possible candidates are further optimized through the first-principles methods as implemented in the widely used VASP code \cite{VASP}. The projector augmented wave methods (PAW)\cite{PAW1,PAW2} and the generalized gradient approximation (GGA) \cite{PBE} are used in this work. A plane wave basis with cutoff energy of 500 eV is used to expand the wave functions for all carbon systems and the Brillouin zone sampling meshes are set to be dense enough to ensure the convergence. All these 2D hydrocarbons are fully optimized until the residual force on each atom is less than 0.01 eV/${\AA}$. The convergence criteria of total energy is set to be 10$^{-6}$ eV and the thickness of the slab-model is set to be larger than 15 ${\AA}$ to avoid spurious interactions between adjacent images. After we identified the ground state configuration, we have systematically investigated its structure, stabilities (dynamical, mechanical and thermal), electronic and mechanical properties. The high-level HSE06 method is also used to confirm the electronic property. To check the dynamical stability, the vibrational spectrum of the ground state configuration is simulated by the open-free PHONOPY code \cite{phonopy} associated with VASP.
\begin{figure}
\begin{center}
\includegraphics[width=\columnwidth]{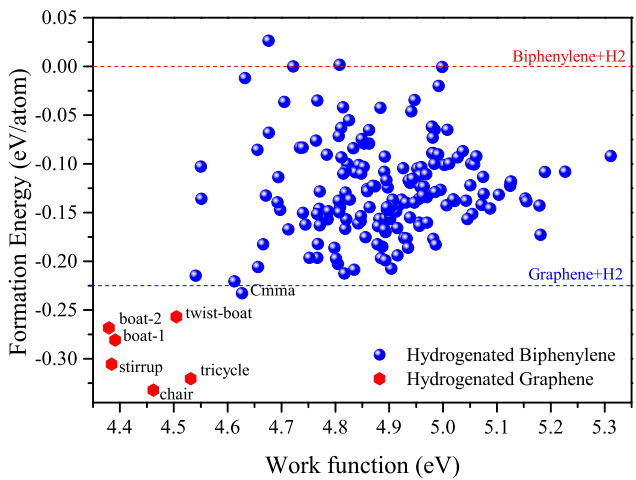}
\caption{The calculated formation energies (refer to Biphenylene sheet and H$_2$) and surface work functions for all possible hydrogenated Biphenylene sheet discovered in this work (blue solid circles) and the well-known graphane allotropes (red solid hexagons).}
\end{center}
\end{figure}

To reveal the energetic stabilities of these 173 new 2D hydrocarbons, we calculated their formation energies as:
\begin{equation}
 \begin{split}
E_f=\frac{E_{tot}-E_H*N_H-E_C*N_C}{N_H+N_C}
 \end{split}
\end{equation}
where E$_{tot}$ is the total energy of the given hydrocarbon system; E$_H$ and E$_C$ are the average energies of reference systems H$_2$ and Biphenylene sheets, respectively. N$_H$ and N$_C$ are the numbers of H-atoms and C-atoms in the given hydrocarbon system, respectively. The formation energies of all the 173 hydrogenated Biphenylene sheets are plotted in Fig. 1 together with their corresponding surface work functions. The six well-known graphane allotropes \cite{prb2003, prb2007, prb2010, prb2011, hcyPSS}(namely, the hydrogenated graphene with chair, tricycle, stirrup, boat-1, boat-2 and twist-boat configurations) are also included in for comparing and the energy of Graphene+H$_2$ is considered as another reference line.

As shown in Fig.1, we can see that most of these hydrogenated Biphenylene sheets possess negative formation energies, which indicate that they can be synthesized from hydrogenating the experimentally synthesized Biphenylene sheet from the energetic views. Among these 173 candidates, the one with Cmma symmetry is identified as the ground state for hydrogenated Biphenylene sheet according to its lowest formation energy of -0.233 eV/atom. Cmma-CH is even more stable than the second reference system of graphene+H$_2$, which indicates that it dose not spontaneously decompose to graphene and H$_2$. Although Cmma-CH is energetically less stable than all the six allotropes for hydrogenating graphene, it dose not means that we have no chance to synthesize Cmma-CH as a new 2D hydrocarbon. We believe that hydrogenation dose not broken the C-C bonds in graphene or Biphenylene sheet to change their fundamental atomic configurations. It just changes the hybridization manners of carbon atoms in the systems and transform the flat sp$^2$-configuration to the buckled sp$^3$-configuration. That is to say, Bipheylene sheet (graphene) can keep its fundamental atomic configuration under hydrogenating and the product is high-probability the Cmma-CH (chair-type graphane) but not the chair-type graphane (Cmma-CH) based on the configuration of graphene (Bipheylene sheet).

We know that hydrogenation of graphene for synthesizing the chair-type graphane is a spontaneous process \cite{S2009}, which means that it is an exothermic reaction. As shown in Fig. 2, the released energies for synthesizing the six most stable graphane allotropes (chair, tricycle, stirrup, boat-1, boat-2 and twist-boat) from hydrogenating graphene can be known as their corresponding negative formation energies (-E$_f$) as defined in equation (1) with graphene and H$_2$ as reference. The released energies (refer to Biphenylene sheet and H$_2$) for synthesizing the six most stable configurations (Cmma, Pmna, Pmma, Pccm, Pma2 and P2/m) of hydrogenated Biphenylene sheet are also shown in Fig. 2. We can see that the released energy in the process of synthesizing Cmma-CH (Pmna, Pmma, Pccm, Pma2 and P2/m ) from hydrogenating Biphenylene sheet is 0.233 eV/atom, which is much remarkable than that of 0.103 eV/atom for synthesizing the chair-type (tricycle, stirrup, boat-1, boat-2 and twist-boat) graphane by hydrogenating graphene. These results indicates that it is more easy to synthesize Cmma-CH (Pmna, Pmma, Pccm, Pma2 and P2/m) than synthesizing the chair-type (tricycle, stirrup, boat-1, boat-2 and twist-boat) graphane by surface hydrogenation.
\begin{figure}
\begin{center}
\includegraphics[width=\columnwidth]{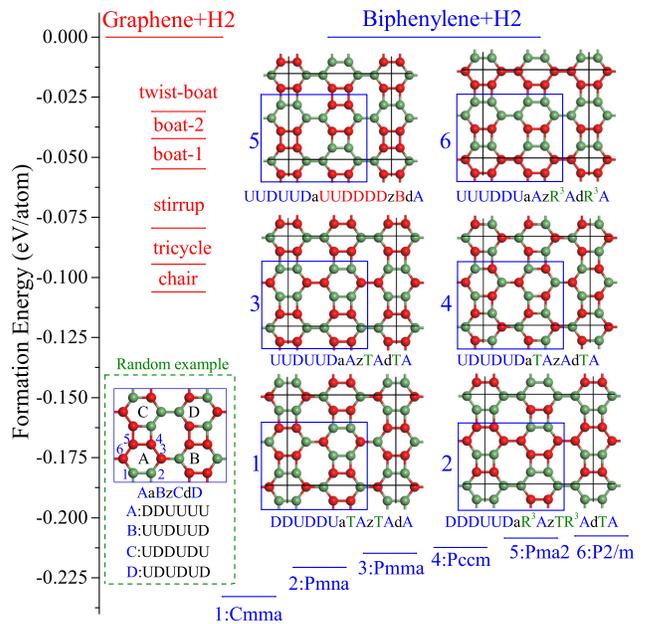}
\caption{The formation energies (refer to graphene and H$_2$) of the six well-known graphane allotropes and the formation energies (refer to Biphenylene sheet and H$_2$) of the six most stable candidates for hydrogenated Biphenylene sheet. The inserted pictures show the up/down sequences in Biphenylene sheet, where red and green balls mean the up and down of carbon atoms, respectively.}
\end{center}
\end{figure}

The structural features of these new 2D hydrocarbons from hydrogenating Biphenylene sheet can be described by the hydrogenation induced up/down sequence of carbon atoms in the systems as inserted in Fig.2. The up/down sequence in hydrogenated graphene (graphane allotropes) have been well-defined before \cite{hcyPSS} and can be studied in Fig.S1. As the random example indicated in Fig.2, the up/down sequence of carbon in each hydrocarbon is recorded as AaBzCdD, in which A, B, C and D are the four benzene rings in the 2$\times$2 supercell of Biphenylene sheet. Interlinks a, z and d mean that B, C and D locate at the armchair, zigzag and diagonal directions of A, respectively. For each benzene ring A, B, C and D, we consider the anticlockwise order of 123456 (starting from the left lower corner) as indicated in Fig. 2 to record the detail up/down sequence, such as A:DDUUUU, B:UUDUUD, C:UDDUDU and D:UDUDUD for the given random example.

The symmetry operators in benzene molecule, namely rotations (R=C$_6$, R123456=612345, R$^2$123456=561234), perpendicular mirrors (m$_{ij}$, exchange i and j, m$_{12}$123456=216543) and horizontal mirror (T, exchange U and D), are used to simplify the record of the up/down sequence. For example, DDUUDUaDDDDUUzUUDDUDdUDDDDU can be simplified as DDUUDUaDDDDUUzTAdRB, where TA means exchange the label U and D in A position for benzene C and RB=RDDDDUU=UDDDDU for benzene D. The up/down sequences for the most stable six hydrogenated Biphenylene sheets are shown in Fig.2 (Cmma, Pmna, Pmma, Pccm, Pma2 and P2/m). We can see that they are exactly different from each other. Tack the the most stable Cmma as example, its sequence is DDUDDUaUUDUUDzUUDUUDdDDUDDU, which can be simplified as DDUDDUaTAzTAdA. It indicate that the hydrogen atoms in Cmma-CH bond to carbon atoms in each benzene with a boat-like (boat-1:DDUDDU) up/down sequence and reversed boat-1 (UUDUUD=TDDUDDU) sequence in adjacent benzene rings. Such structural features of these hydrogenated Biphenylene sheets can also be known from their optimized crystalline structures as shown in the supplementary Fig. S2.

The energy difference between the most stable Cmma (-0.233 eV/atom) configuration and the second stable Pmna (-0.221 eV/atom) one is 12 meV/atom, which indicates we have higher probability to synthesize Cmma in the process of hydrogenating Biphenylene sheet rather than Pmna and other phases (Pmma, Pccm, Pma2 and P2/m). The optimized crystal structures of Cmma (including top/side views) in both primitive and crystalline cells are plotted in Fig.3 (a) and (b), respectively. The ground state configuration Cmma contains four benzene molecules in its crystalline cell and all of them are inequivalent due to its Cmma symmetry. The optimized lattice constants for such a crystalline cell are a=8.913 {\AA}, b=7.611 {\AA} and c=20.00 {\AA}. There are 24 H-atoms and 24 C-atoms in the crystalline structure of Cmma and only 2 H-atoms (H1:0.881 0.811 0.428 and H2: 0.602 0.500 0.438) and 2 C-atoms (C1:0.838 0.835 0.478 and C2:0.586 0.500 0.492) are inequivalent. The average length of C-C bonds is 1.55 {\AA}, which is slightly larger than those in Biphenylene sheet. The average length of C-H bonds is 1.09 {\AA}, which is very similar to those in graphane allotropes.

\begin{figure}
\begin{center}
\includegraphics[width=\columnwidth]{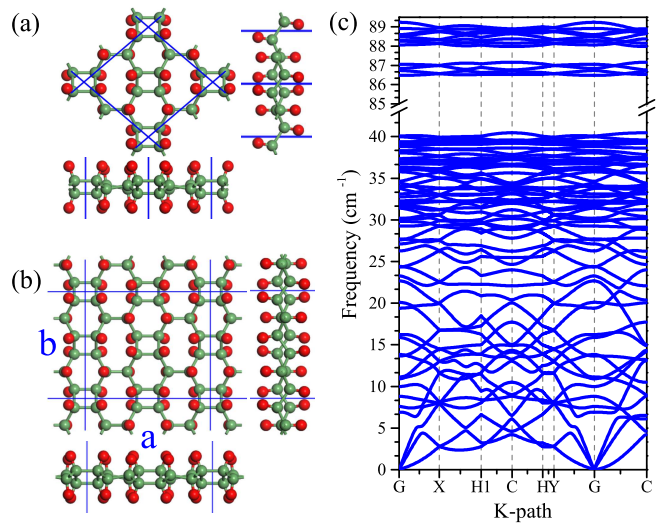}
\caption{The optimized primitive (a) and crystalline (b) cells of Cmma-CH in both top and side views, where green and red balls represent carbon and hydrogen atoms. The calculated phonon spectrum (c) of Cmma-CH based on a 2$\times$2 supercell of its primitive cell.}
\end{center}
\end{figure}
To confirm if Cmma-CH can keep its crystalline structure well under small vibrations, we calculated its vibrational spectrum as shown in Fig.3 (c) based on a 2$\times$2 supercell (containing 96 atoms) of its primitive cell. It is clearly that there is no any negative frequency in the simulated phonon band structure. We have also checked its phonon density of state in the whole Brillouin Zone and found no any imaginary modes. These results suggest that Cmma-CH is a dynamically stable phase for 2D hydrocarbon, which dose not spontaneously transform to any other hydrocarbon phases like chair-type graphane. The simulated ab initio molecular dynamics (MD) for Cmma-CH (based on the same supercell used in phonon calculation) at both 300 K and 500 K in 5 ps (with time step of 1 fs) are shown in Fig. S3. The inserted snapshots suggest that Cmma-CH can maintain its fundamental crystal structure well with just small thermal oscillations in the MD simulation processes. The oscillations in total energies near the ground states as shown in Fig. S3 are reasonable and predictable, and it suggests that Cmma-CH is thermally stable at room (300 K) and even higher temperature (500 K).
\begin{figure}
\begin{center}
\includegraphics[width=\columnwidth]{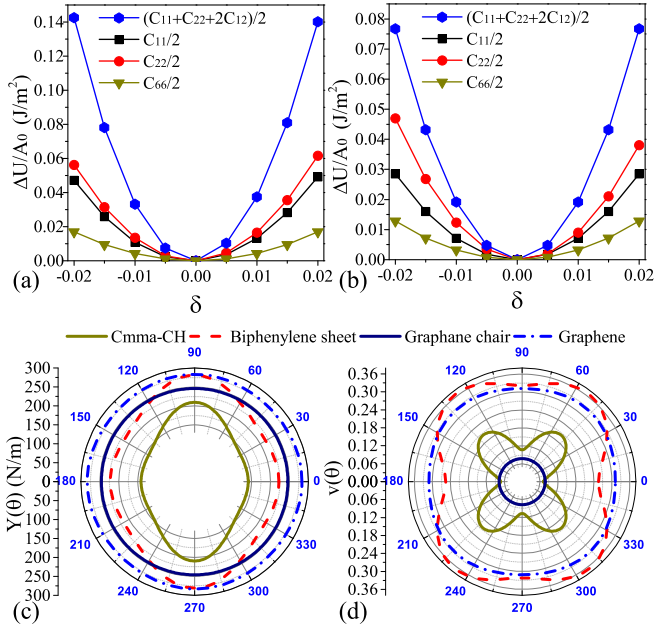}
\caption{The calculated deformation energies as functions of external strains for Biphenylene sheet (a) and Cmma-CH (b). The direction-dependent Young's modulus Y($\theta$) (c) and Poisson's ratio v($\theta$) (d) of Biphenylene sheet and Cmma-CH together with those for graphene and chair type graphane.}
\end{center}
\end{figure}

To obtain the elastic constants C$_{11}$, C$_{22}$, C$_{12}$ and C$_{66}$ of Cmma-CH (Biphenylene sheet) to check its elastic stability and evaluate its mechanical property, we fit its crystalline-cell energy U(A$_0$, $\varepsilon$) with the state of equation as function of 2D strains $\varepsilon$$_{11}$ (zigzag direction), $\varepsilon$$_{22}$ (armchair direction) and $\varepsilon$$_{12}$ (shear strains) as below:
\begin{equation}
 \begin{split}
&\frac{\Delta U}{A_0}=\frac{C_{11}\varepsilon^2_{11}}{2}+\frac{C_{22}\varepsilon^2_{22}}{2}+C_{12}\varepsilon_{11}\varepsilon_{22}+\frac{C_{66}\varepsilon^2_{12}}{2}
 \end{split}
\end{equation}
where $\Delta$U=U(A$_0$, $\varepsilon$$_{11}$, $\varepsilon$$_{22}$, $\varepsilon$$_{12}$)-U(A$_0$, $\varepsilon$$_{11}$=0, $\varepsilon$$_{22}$=0, $\varepsilon$$_{12}$=0) is the deformation energy and A$_0$ is the corresponding crystalline-cell area at equilibrium state. Four groups of independent external stains ($\varepsilon$$_{11}$, $\varepsilon$$_{22}$, $\varepsilon$$_{33}$, $\varepsilon$$_{23}$, $\varepsilon$$_{13}$, $\varepsilon$$_{12}$) are applied to the system for calculating the deformation energies $\Delta$U and four different equations related to the deformation energies can be used to solve the elastic constants (C$_{11}$, C$_{22}$, C$_{12}$ and C$_{66}$) as below:
\begin{equation}
 \begin{split}
&\varepsilon_{11}=\delta, \varepsilon_{ij}=0:\Delta U=\frac{A_0}{2}C_{11}\delta^2
 \end{split}
\end{equation}
\begin{equation}
 \begin{split}
&\varepsilon_{22}=\delta, \varepsilon_{ij}=0:\Delta U=\frac{A_0}{2}C_{22}\delta^2
 \end{split}
\end{equation}
\begin{equation}
 \begin{split}
&\varepsilon_{11}=\varepsilon_{22}=\delta, \varepsilon_{ij}=0:\Delta U=\frac{A_0}{2}(C_{11}+C_{22}+2C_{12})\delta^2
 \end{split}
\end{equation}
\begin{equation}
 \begin{split}
&\varepsilon_{12}=\delta, \varepsilon_{ij}=0:\Delta U=\frac{A_0}{2}C_{66}\delta^2
 \end{split}
\end{equation}

The calculated deformation energies as function of external strains ($\delta$=-0.02 -0.01 0 0.01 0.02) for Biphenylene sheet and Cmma-CH determined by equation (2)-(6) are shown in Fig. 4 (a) and (b). Based on a quadratic fitting of the results shown in Fig. 4, the elastic constants of Biphenylene sheet are calculated to be C$_{11}$=242.293 N/m, C$_{22}$=305.883 N/m, C$_{12}$=77.967 N/m and C$_{66}$=81.273 N/m as listed in Table I together with those for graphene, Cmma-C and the chair-type graphane. These elastic constants for Biphenylene sheet are slightly small than those of graphene (C$_{11}$=313.307 N/m, C$_{22}$=313.307 N/m, C$_{12}$=97.472 N/m and C$_{66}$=104.523 N/m). After hydrogenation, these elastic constants are obviously reduced to be C$_{11}$=142.351 N/m, C$_{22}$=210.865 N/m, C$_{12}$=15.148 N/m and C$_{66}$=61.532 N/m, respectively, which are also smaller than those for the chair-type graphane (C$_{11}$=247.175 N/m, C$_{22}$=247.175 N/m, C$_{12}$=19.035 N/m and C$_{66}$=114.193 N/m). These elastic constants satisfy the mechanical stability criteria (C$_{11}$$\times$C$_{22}$-C$^2_{12}$$>$0, C$_{66}$$>$0) for 2D materials \cite{judge}, suggesting that both Biphenylene sheet and Cmma-CH are mechanically stable to resist small deformations.

Based on the calculated independent elastic constants, the direction-dependent Young's modulus and Poisson's ratio for Biphenylene sheet and Cmma-CH can be evaluated as below formulas for 2D systems\cite{Yandv}:
\begin{equation}
 \begin{split}
&Y(\theta)=\frac{C_{11}C_{22}-C_{12}^2}{C_{11}s^4+C_{22}c^4+(\frac{C_{11}C_{22}-C_{12}^2}{C_{66}}-2C_{12})s^2c^2}
 \end{split}
\end{equation}

\begin{equation}
 \begin{split}
&v(\theta)=-\frac{(C_{11}+C_{22}-\frac{C_{11}C_{22}-C_{12}^2}{C_{66}})s^2c^2-C_{12}s^4c^4}{C_{11}s^4+C_{22}c^4+(\frac{C_{11}C_{22}-C_{12}^2}{C_{66}}-2C_{12})s^2c^2}
 \end{split}
\end{equation}
where $\theta$ is the direction angle relative to the positive a-axis (armchair direction) as indicated in the crystalline cell in Fig.3 (b). The letters s and c in the equations (7) and (8) are used for representing the values of sin($\theta$) and cos($\theta$), respectively. The calculated Young's modulus and Poisson's ratio, for Biphenylene sheet, Cmma-CH, graphene as well as the chair-type graphane, are plotted in Fig. 4 (c) and (d), respectively. It is clearly that Biphenylene sheet and Cmma-CH with orthorhombic symmetry present more obvious anisotropy in both Young's modulus and Poisson's ratio in comparing with the hexagonal graphane and graphene with higher symmetry. And we can see that surface hydrogenation will obviously weaken the mechanical properties of both Biphenylene and graphene, reducing the Young's modulus, the Poisson's ratio, as well as the elastic constants as summarized in Table I.
\begin{table}
\center
\caption{The calculated independent elastic constants (C$_{11}$, C$_{22}$, C$_{12}$ and C$_{66}$ in unit of N/m) for graphene, graphane, Biphenylene sheet and Cmma-CH, as well as the corresponding Young's modulus and Poisson's ratio in the armchair (Y(0) and v(0) in unit of N/m) and zigzag direction Y(90) and v(90).}
\begin{tabular}{c c c c c c c c c c }
\hline \hline
&System &Graphene  &Graphane &Biphenylene sheet &Cmma-CH \\
\hline
&C$_{11}$  &313.307 &247.175 &242.293  &142.351    \\
&C$_{22}$  &313.307 &247.175 &305.883  &210.865    \\
&C$_{12}$  &97.472  &19.035  &77.967   &15.148    \\
&C$_{66}$  &104.523 &114.193 &81.273   &61.532    \\
&Y(0)      &282.983 &245.709 &222.421  &141.263    \\
&Y(90)     &282.983 &245.709 &280.794  &209.235   \\
&v(0)      &0.311   &0.077   &0.255    &0.072    \\
&v(90)     &0.311   &0.077   &0.322    &0.106    \\
\hline \hline
\end{tabular}
\label{tab1}
\end{table}

Previous literatures have shown that hydrogenation can transform the semi-metallic graphene to insulating graphane \cite{prb2007, prb2010, prb2011, hcyPSS} by changing the hybridizing manners of carbon atoms in graphene from sp$^2$ to sp$^3$. In this work, our results show that hydrogenation can also induce same transitions in both structure and electronic property in Biphenylene sheet. As shown in Fig. 5 (a), the electronic band structures of Biphenylene sheet from both PBE and HSE06 calculations suggest that it is a just normal metal. After hydrogenation, all the sp$^2$-hybridized states in Biphenylene sheet are changed to sp$^3$ and the planner system transform to bulked Cmma-CH with insulating property as its band structures shown in Fig. 5 (b). The HSE06 (PBE) based results show that Cmma-CH is an indirect band gap insulator with a band gap of 4.645 eV (3.631 eV), which is slightly larger than that of the chair-type graphane (HSE06: 4.437 eV and 3.486 eV). And such a band gap can be effectively modulated by the external uniaxial ($\varepsilon$$_{11}$=$\delta$ or $\varepsilon$$_{22}$=$\delta$) and biaxial ($\varepsilon$$_{11}$=$\varepsilon$$_{22}$=$\delta$) strains as the results calculated based on PBE in Fig. 5 (c).
\begin{figure}
\begin{center}
\includegraphics[width=\columnwidth]{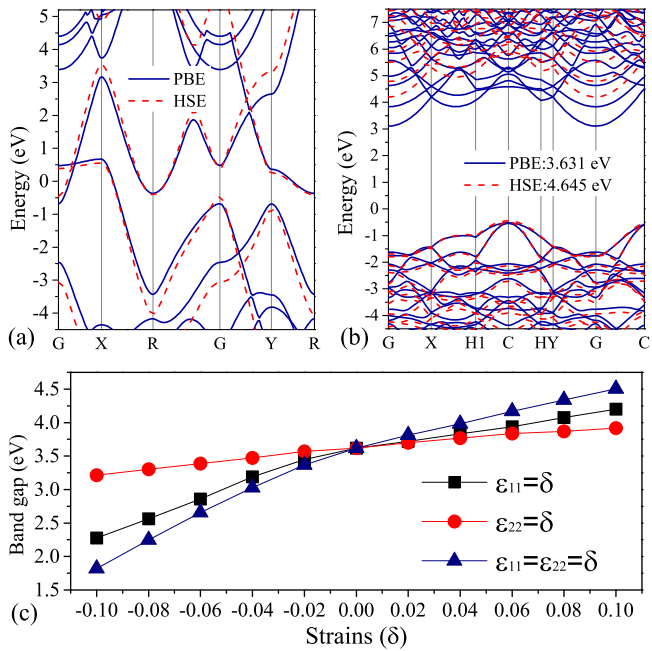}
\caption{The calculated electronic band structures for Biphenylene sheet (a) and Cmma-CH (b) from both PBE and HSE06 calculations. The modulating effects of external strains on the band gap of Cmma-CH based on PBE calculation (c).}
\end{center}
\end{figure}

In summary, the ground state of hydrogenated Biphenylene sheet is confirmed to be the Cmma structure containing four benzene rings in its crystalline cell with a Cmma symmetry. The hydrogen atoms in Cmma-CH bond to carbon atoms in each benzene ring with a boat-1 (DDUDDU) sequence and a reversed boat-1 (UUDUUD) sequence in adjacent benzene rings. Cmma-CH is energetically less stable than the experimentally synthesized chair-type graphane from hydrogenating graphene. However, its formation energy from hydrogenating Biphenylene sheet is remarkably lower than that of chair-type graphane from hydrogenating graphene, which indicates that it can be potentially synthesized from hydrogenating Biphenylene sheet. We have also confirmed that such a new 2D hydrocarbon phase is a mechanically and dynamically stable insulator with an indirect band gap of 4.645 eV based on HSE06 method. These results in our work show that surface hydrogenation can effectively modulate the structural, mechanical and electronic properties of the experimentally synthesized Biphenylene sheet and it is an promising approach to create a new 2D hydrocarbon material that has never be discovered in the nature.

This work is supported by the National Natural Science Foundation of China (Grants No. 11974300 and 11974299) and the Program for Changjiang Scholars and Innovative Research Team in University (No. IRT13093).
\bibliographystyle{apsrev}

\end{document}